# Relationship between Loss of Load Expectation and Frequency of Load Loss Events

**Chanan Singh**
**University Distinguished Professor**
**Electrical and Computer Engineering Department**
**Texas A&M University**

**September 15, 2026**

## Relationship between Loss of Load Expectation and Frequency of Load Loss Events

**Chanan Singh,**
**University Distinguished Professor**
**Electrical and Computer Engineering Department**
**Texas A&M University**

**Abstract:**
Loss of Load Expectation (LOLE), originally calculated as the expected number of days of load loss, has also been used as a frequency metric. This paper shows that LOLE can indeed be calculated as a frequency metric, but only with restricted conditions, and that it differs fundamentally from the standard frequency metric employed in the frequency and duration (F&D) approach. A lack of understanding of these conditions can lead to incorrect applications and interpretations.

**Introduction**
In power system reliability and resource adequacy literature, Loss of Load Expectation (LOLE) is technically defined as the expected number of days or hours of loss of load. However, several major grid operator reports, energy research papers, and technical standards informally refer to LOLE as a "frequency" metric, or use it synonymously with the frequency of outage events. Below are representative examples in which LOLE is characterized or explicitly called a measure of frequency.

1. Energy Systems Integration Group (ESIG) Reports
In studies assessing modern grids, ESIG frequently simplifies the explanation of standard resource adequacy metrics by categorizing LOLE as the proxy for frequency. It says, *"The most common resource adequacy criterion today is the one-day-in-10-year loss-of-load expectation (LOLE) in North America. However, this single metric only measures the frequency of outages and doesn't capture the size, duration, or timing of generation shortfalls."* [1]
2. Midcontinent Independent System Operator (MISO) Planning Documents
During regulatory and stakeholder meetings regarding Resource Adequacy Risk Metrics, grid planners align metrics with specific risk dimensions (Frequency vs. Size vs. Duration). [2] In evaluating metric attributes, MISO explicitly expresses:

- *"Loss of load expectation (LOLE): to capture frequency of events."*
- *"Expected unserved energy (EUE): to capture size of events."*
- *"Loss of load hours (LOLH): to capture event duration."*

3. North American Electric Reliability Corporation (NERC) Historical & Technical Frameworks [3]
Because the standard target of "1 day in 10 years" (0.1 days/year) is conceptually framed by the public as "one event every ten years," historical NERC definitions occasionally bridge the gap by linking LOLE directly to event occurrences.
The text explains that planning *"has historically been based strictly on Loss of Load Expectation (LOLE), or the number of firm load shed events an electric system expects over a period of one or more years... the primary economic consequence of reliability-related events is not necessarily in the frequency or duration of firm load shed events [LOLE]..."*

These examples indicate some confusion regarding the various indices proposed in the literature [4,5,6]. This paper aims to clarify the differences between these indices and to examine whether it is correct to refer to LOLE as a frequency. Finally, it shows that LOLE is indeed a frequency metric under certain specified conditions, but that it should not be confused with the standard frequency and duration indices.

**Indices or Metrics for Resource Adequacy**

The various indices proposed for resource adequacy are briefly explained below.

1. Loss of Load Expectation (LOLE)
   Loss of load expectation has two forms, based on daily peaks or hourly loads.

- DLOLE or LOLE is the expected number of days per year on which insufficient generating capacity is available to serve the daily peak load. Now some utilities have replaced the daily peak by load loss during any time of the day.
- HLOLE or LOLH is the expected number of hours per year when insufficient generating capacity is available to serve the load.
- The LOLE index gives no information on a number of important system reliability attributes:
    - Magnitude of capacity shortages when they occur.
    - Duration of capacity shortage events. It should be pointed out that HLOLE gives the mean hours of loss of load in a year but not the mean duration of load loss events.
    - Expected amount of unserved energy

2. Frequency and Duration of Capacity Shortage Events (F&D)
    - Frequency of generating capacity shortage events is defined to be the expected (average) number of such events per year.
    - Duration is the expected length of capacity shortage periods when they occur.
    - F&D indices use hourly load information and thus reflect the influences of daily load cycle shape.
    - F&D methods model unit parameters more fully than those models used in LOLE. These indices need the transition rate between various states of generators whereas the LOLE and EUE indices need only the probabilities of various states of generators.
    - F&D indices are conceptually superior to LOLE. They have, however, greater data requirements.
    - It should be noted that LOLH is the product of the frequency and mean duration.

3. Expected Unserved Energy (EUE)
    - The EUE index measures the expected amount of energy which will fail to be supplied per year due to generative capacity differences and/or shortages in basic energy supplies.

**Computation of Frequency of Load Loss**

The relationship between load and available capacity is shown in Figure 1. This figure conceptually illustrates the events of capacity deficiency or load loss: how often they occur, the duration of the events when they occur, and the magnitude of load loss during an event. It should be noted that whereas LOLH is the mean value of load loss duration over the whole year, the mean duration in the F&D indices is the average load loss duration *per event*.

To understand the frequency index, one must understand how a load loss event can occur. It can happen either because of a change in generation or a change in load. For example, the system can transition into load loss because the available capacity decreases due to generation failure, or because the load increases beyond available capacity. Thus, the frequency of load loss has two components : one originating from generation and the other from load variation. For computing the component due to generation, information on the transition rates between generation states is required, since:

$$\text{Frequency of transition from state } i \text{ to } j = P(i) \times \lambda_{i \to j} \qquad (1)$$

where $P(i)$ is the probability of state $i$ and $\lambda_{i \to j}$ is the transition rate from state $i$ to state $j$.

For the component due to load, information on the hourly variation of load is required. This hourly load variation can then be converted into the frequency of load transitions from one state to another. The two pieces of

information—generation variation and load variation—can then be combined to give the frequency of load loss events.

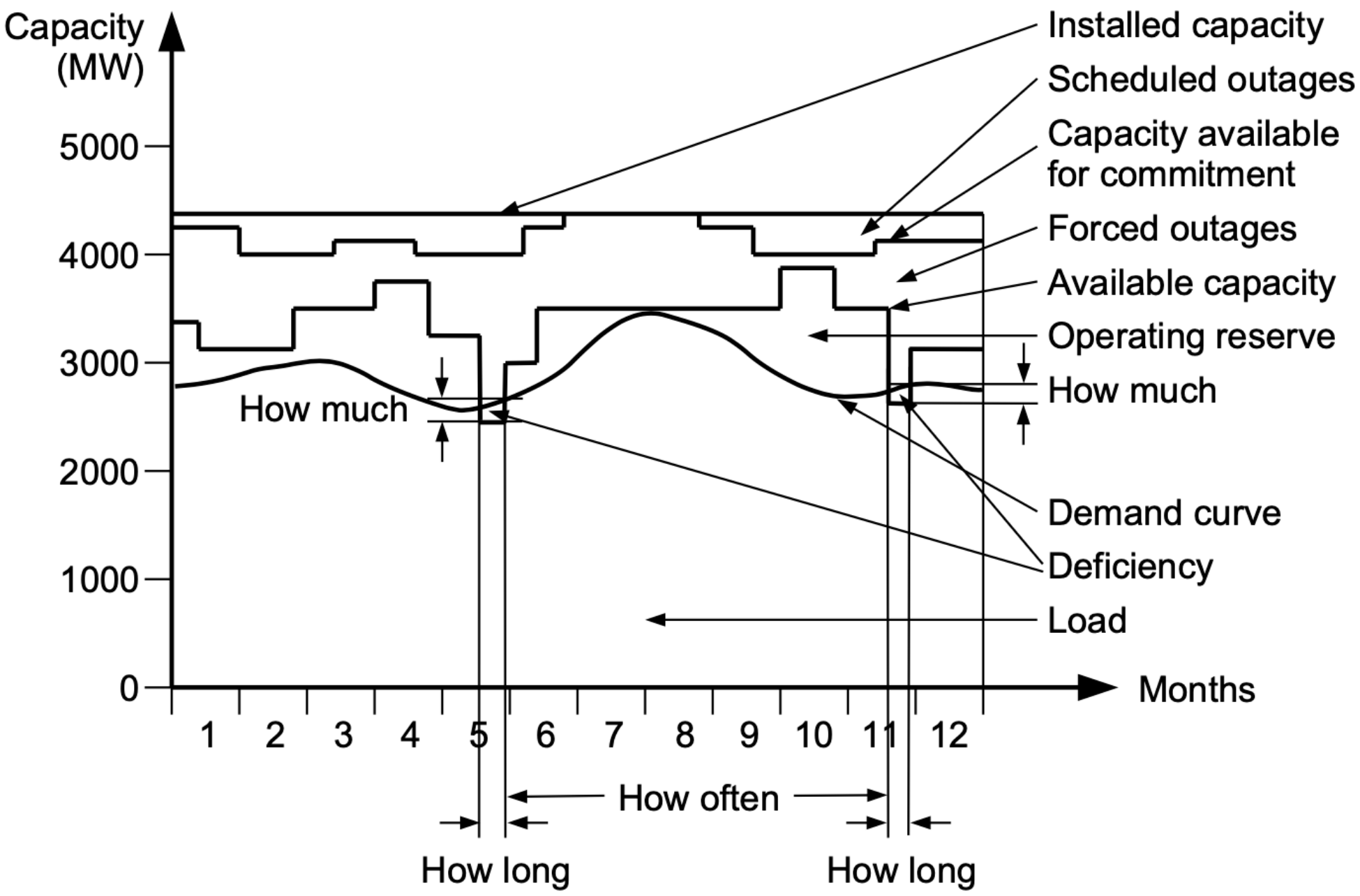

Fig 1. Relationship between generation and load [4]

**Why is LOLE being called a frequency metric?**

Historically, LOLE was calculated from the Loss of Load Probability (LOLP). LOLP was obtained by convolving the capacity outage probability distribution with the probability distribution of daily peak loads and represents the probability that the daily peak exceeds the available generation. LOLE was then calculated as an expected number of days by multiplying LOLP by 365 (the number of days in the year), and was thus termed the expected number of days on which the peak load is not satisfied by the available generation. The "one day in 10 years" target value translates to:

$$\text{LOLP} = \frac{0.1}{365} = 0.00027397 \qquad (2)$$

The LOLE value of 0.1 has commonly been referred to as "one day in 10 years." When stated this way, it does sound like a frequency metric, although it is actually calculated as an expectation. We show later that LOLE can indeed be calculated or interpreted as a frequency under restricted conditions.

Another way to view LOLE is through Monte Carlo simulation. Suppose we perform a sequential Monte Carlo simulation with $N$ yearly replications of generation versus load and observe loss of peak load on $n$ days out of the $N$ years. Then the estimate of LOLE is:

$$\text{LOLE} = \frac{n}{N} \qquad (3)$$

From this perspective, calling LOLE a frequency metric resembles the frequency interpretation of probability, in which probability is found by dividing the number of favorable trials by the total number of trials. This metric, however, is *not* the same as the complete frequency index in the F&D approach

**LOLH, LOLE and EUE**

These three metrics can be calculated from the probability distributions of generators and load. Hourly load is needed for LOLH and EUE and only daily peaks are needed for LOLE. These indices are therefore probability-based indices. It must be emphasized that LOLH gives the total duration of load loss over the interval.

**F&D Indices**

For the frequency and duration indices as known in the literature, we require the interstate transition rates of the generators and the hourly load. These indices provide the frequency of load loss and the mean duration of an event.

Consider a system in which generation has five states and load has three states, as shown in Figure 2. The generation states range from 200 MW to 0, and the load has three states ranging from 0 to 100 MW. The states of loss of load (negative margin) are shown below the marked boundary. Consider a loss of load state with a capacity of 50 MW and a load of 100 MW. The system can move out of this loss of load state in two ways:

1. The load remains at 100 MW, but the capacity increases from 50 MW to a higher level. This contributes to the frequency due to generation change, for which the generation interstate transition rates are required.
2. The capacity remains at 50 MW, but the load decreases from 100 MW to a lower state. This is the contribution to frequency due to load variation. This component can be calculated as:

$$P(\text{Cap} = 50) \times P(\text{Load} = 100) \times \lambda_{\text{load},100\rightarrow\text{lower}} \qquad (4)$$

It is worth noting that the generation interstate transition rates are not needed for the frequency component due to load. The frequency of load transitions at a given level can be calculated by scanning the load sequentially and counting the number of times the load moves from below the level to above it (see Figure 3).

| Load → Capacity ↓ | 100 | 50 | 0 | |
|---|---|---|---|---|
| 200 | 100 | 150 | 200 | |
| 150 | 50 | 100 | 150 | ← Margins |
| 100 | 0 | 50 | 100 | = Cap- Load |
| 50 | -50 | 0 | 50 | |
| 0 | -100 | -50 | 0 | |

Margin ≤ -50MW

Figure 2. Interaction between generation and load.

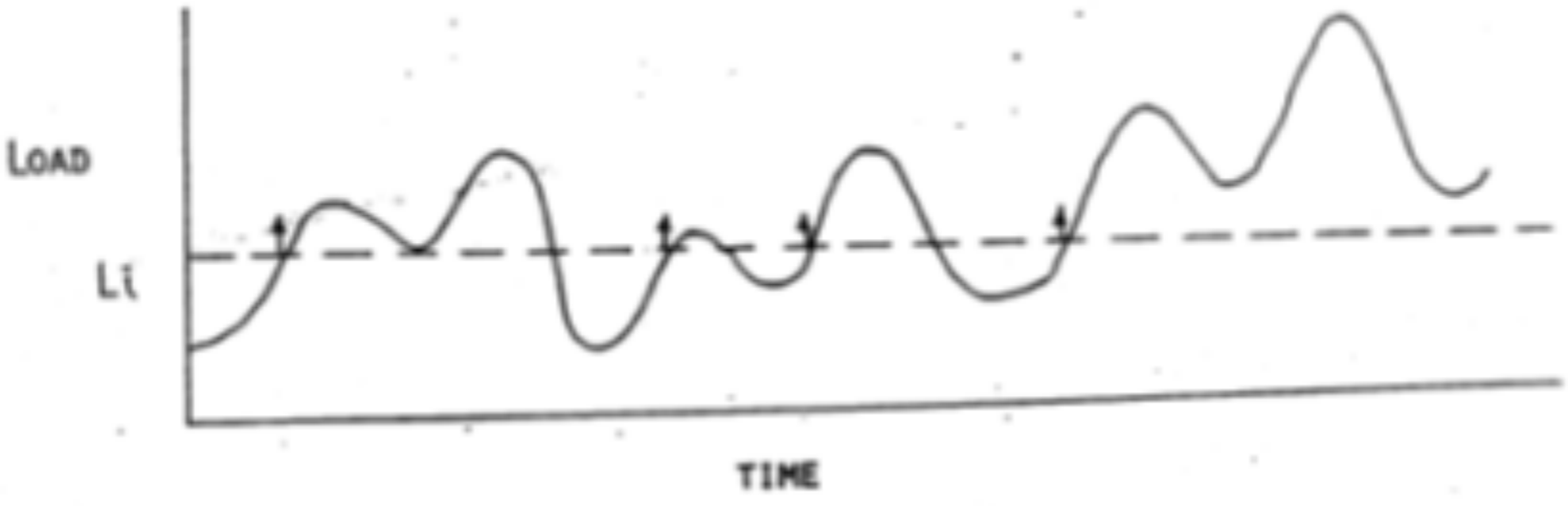

Figure 3. Counting Frequency Events

**Relationship between LOLE and Frequency**

Assume that we have calculated the probabilities of the various capacity levels of generation, and that we can find the cumulative probability distribution of peak loads (i.e., the probabilities of the peak load exceeding a given level). Then loss of load probability:

$$\text{LOLP} = \sum_i P[C(i)] \cdot P[PL > C(i)] \qquad (5)$$

and

$$\text{LOLE} = \text{LOLP xTP} \qquad (6)$$

Now,

$$P[PL > C(i)] = \frac{NP > C(i)}{TP} \qquad (7)$$

where $NP$ is the number of daily peaks exceeding $C(i)$, and $TP$ is the total number of daily peaks in the interval (365 in a year). Substituting (7) into (6):

$$\text{LOLE} = \sum_i P[C(i)] \cdot \frac{NP > C(i)}{TP} \quad TP \qquad (8)$$

$$= \sum_i P[C(i)] \times [NP > C(i)] \qquad (9)$$

=Expected Loss of daily peaks

= Frequency of loss of peak (10)

Equation (10) shows that LOLE equals the frequency of load loss under the following conditions:

1. The load model considers only load loss at daily peaks.
2. This frequency does not include the component due to generation changes.
3. This frequency does not include the variation of hourly loads.
4. This frequency is not the same as the standard frequency index, and cannot be used to calculate the mean duration of load loss events.

**Conclusions**

LOLE was original proposed as an expectation, expected number days of load loss at the peak. Some publications present this index as a frequency of load loss days per year. This paper shows that LOLE can indeed be calculated as a frequency but his frequency is different than the frequency of load loss in the standard frequency and duration approach. Examples of difference are:

1. In F&D approach the main indices are probability, frequency and mean duration of load loss. In LOLE, the concept of mean duration does not make any sense.
2. In F&D, frequency has components of generation changes and load changes but in LOLE, only component is peak load changes.
3. In F&D approach product of frequency and mean duration gives LOLH, ie, expected number of hours of load loss. This does not hold when LOLE is used as a frequency.